\documentclass[12pt]{article}
\usepackage{hyperref}
\usepackage{cite}
\usepackage{xcolor}
\usepackage{graphicx}
\usepackage{caption,subcaption}
\usepackage{amsmath}
\usepackage{amssymb}
\usepackage{xspace}
\usepackage{verbatim}
\usepackage{mathtools}
\usepackage{tikz-cd}
\usepackage{braket}
\usepackage{url}
\usepackage[normalem]{ulem}

\makeatletter
\@addtoreset{equation}{section}

\makeatletter
\renewcommand\section{\@startsection {section}{1}{\z@}%
                                   {-3.5ex \@plus -1ex \@minus -.2ex}
                                   {2.3ex \@plus.2ex}%
                                   {\normalfont\large\bfseries}}
\renewcommand\subsection{\@startsection{subsection}{2}{\z@}%
                                     {-3.25ex\@plus -1ex \@minus -.2ex}%
                                     {1.5ex \@plus .2ex}%
                                     {\normalfont\bfseries}}

\def\baselinestretch{1.2}
\newcommand{\be}{\begin{equation}}
\newcommand{\ee}{\end{equation}}
\newcommand{\beq}{\begin{eqnarray}}
\newcommand{\eeq}{\end{eqnarray}}

\newcommand{\tr}{{\rm Tr}}
\newcommand{\gone}[1]{{}}

\definecolor{amber}{rgb}{1.0, 0.75, 0.0}

\begin{document}
\begin{titlepage}
\begin{flushright}
MAD-TH-19-03
\end{flushright}

\vfil

\begin{center}

{\bf \Large
Weak Gravity Conjecture, Black Hole Entropy, and Modular Invariance
}

\vfil

Lars Aalsma$^{a}$, Alex Cole$^{b}$, and Gary Shiu$^{b}$

\vfil

l.aalsma@uva.nl, acole4@wisc.edu, shiu@physics.wisc.edu

$^a$Institute for Theoretical Physics and Delta Institute for Theoretical Physics\\ University of Amsterdam, PO Box 94485, 1090 GL Amsterdam, The Netherlands\\
$^b$Department of Physics, University of Wisconsin, Madison, WI 53706, USA

\vfil
\end{center}

\begin{abstract}

\noindent In recent literature, it has been argued that a mild form of the Weak Gravity Conjecture (WGC) is satisfied by wide classes of
effective field theories in which higher-derivative corrections can be shown to shift the charge-to-mass ratios of extremal black holes to larger values. However, this mild form 
does not directly constrain low-energy physics because the black holes satisfying the WGC have masses above the cutoff of the effective theory. In this note, we point out that in string theory modular invariance can connect a light superextremal state to heavy superextremal states; the latter collapse into black holes at small string coupling. In the context of heterotic string theory, we show that these states are black holes that have $\alpha'$-exact charge-to-mass ratios exceeding the classical extremality bound. This suggests that modular invariance of the string partition function can be used to relate the existence of a light superextremal particle to the positive shift in charge-to-mass ratio of extremal black holes.
\end{abstract}
\vspace{0.5in}

\end{titlepage}
\renewcommand{\baselinestretch}{1.05}  


\section{Introduction}
Understanding the low-energy predictions of quantum gravity is a task of utmost importance. From a naive effective field theory perspective, quantum gravity effects might only appear near the Planck scale, casting doubt on our ability to observe them. Luckily, nature has been more generous to us in that quantum gravity appears not to decouple completely from IR physics. This notion has been captured by the idea of the ``swampland'' \cite{Vafa:2005ui,Ooguri:2006in}
(see \cite{Brennan:2017rbf,Palti:2019pca} for reviews). The swampland program states that quantum gravity imposes certain consistency conditions on its low-energy effective theories; we can view these as universal predictions of quantum gravity. One such condition is the Weak Gravity Conjecture (WGC) \cite{ArkaniHamed:2006dz}, which requires that there exists a charged (super)extremal state allowing (nonsupersymmetric) extremal black holes to decay.

Even if the WGC is true in general, its implications for a low-energy observer are only clear when we take a sharply-defined statement of the conjecture. In terms of the ``electric'' WGC, 
some strong statements have been proposed. These include requiring
the existence of a \emph{light} charged particle \cite{ArkaniHamed:2006dz} and a tower of states \cite{Montero:2016tif,Heidenreich:2016aqi,Andriolo:2018lvp}. 
On the other hand, the mild form of the WGC only requires some state into which an extremal black hole can decay, which can be another black hole \cite{Kats:2006xp,Cheung:2018cwt,Hamada:2018dde,Bellazzini:2019xts}. In this case, higher-derivative corrections shift the charge-to-mass ratio ${\cal Z}\equiv Q/M$ ($\equiv1$ as $M\to \infty$) of large extremal black holes positively so that larger extremal black holes can decay into them. 
The restriction to large black holes allows one to include only the leading higher-derivative correction,
but at the cost of weakening the constraints on low-energy physics.

The difference between the mild and strong forms of the WGC is particularly important in the context of axion inflation, where (in the axionic statement of the WGC) the strong form places tighter constraints on transplanckian axion field ranges \cite{Rudelius:2015xta,Montero:2015ofa,Brown:2015iha,Brown:2015lia}. 

In this note, we suggest that having a light superextremal particle in the spectrum is closely tied to a positive shift in charge-to-mass ratio of an extremal black hole, effectively relating the mild and strong form of the WGC. 
We illustrate this with perturbative string theory, but because the essence of our argument relies on universal modular transformation properties of partition functions our results may be applicable more generally such as to the non-perturbative F-theory setup recently considered in \cite{Lee:2018urn,Lee:2018spm,Lee:2019xtm}.
We show that modular invariance imposes certain constraints on the perturbative spectrum, allowing us to relate light superextremal perturbative string states to superextremal heavy states. These heavy states collapse to form black holes at small string coupling and connect the strong and mild 
forms of the WGC. We use an $\alpha'$-exact entropy matching to show that in the context of the heterotic string one can identify ${\cal Z}>1$ higher-derivative corrected black holes as superextremal perturbative string states at  
non-zero coupling.

The rest of this note is organized as follows. In Section \ref{sec:HD} we briefly review how higher-derivative corrections can modify the black hole charge-to-mass ratio. In Section \ref{sec:MI} we show how modular invariance relates light and heavy perturbative string states and their charge-to-mass ratios. In Section \ref{sec:BHS} we identify certain extremal heterotic black holes as superextremal fundamental string states via an $\alpha'$-exact entropy matching. We briefly remark in Section \ref{sec:EC} how our arguments might be related to the positivity of black hole entropy corrections. We discuss our results in Section \ref{sec:DIS}.
\section{Higher-Derivative Corrections to Extremal Black Holes} \label{sec:HD}
Although general relativity is not renormalizable, we can still consider it as an effective theory. General relativity should then be modified by higher-derivative corrections that vanish in the large distance/small curvature limit. In string theory, these corrections are controlled by $(\alpha')^n \mathcal{R}_n$, where ${\cal R}_n$ is a curvature invariant with $2(n+1)$ derivatives, which should be small relative to the scalar curvature $\mathcal{R}$ for a well-defined perturbative expansion. These $\alpha'$ corrections modify the solutions to the theory and in general change the black hole extremality bound, shifting the charge-to-mass ratio $\mathcal{Z}_{\textrm{ext},\alpha'}$ above which a black hole forms a naked singularity. The modified ratio is a function of $M$, asymptoting to the uncorrected bound in the infinite mass limit, $\lim_{M\to \infty}\mathcal{Z}_{\textrm{ext},\alpha'}=1$, where $\alpha'$ corrections for typical (large) black holes become negligible.
\begin{figure}\centering
\includegraphics[width=0.4\textwidth]{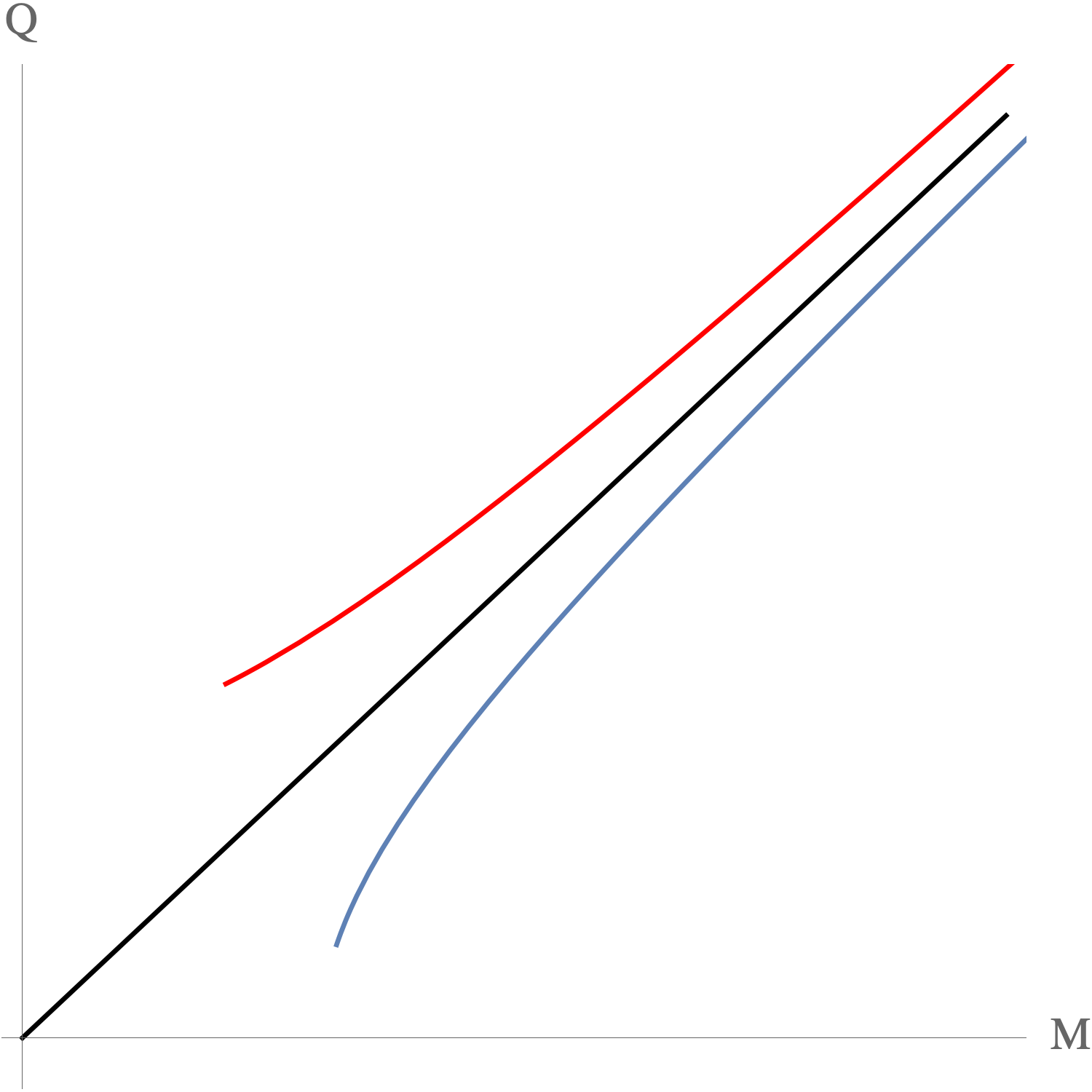}
\caption{Consistency with the decay of extremal black holes dictates that higher-derivative corrections should shift an extremal black hole's charge-to-mass ratio positively (red curve). The blue curve would imply an infinite family of stable nonsupersymmetric states.}\label{fig:HD}
\end{figure}
Noting that the mild form of the WGC requires states with $\mathcal{Z}>1$ and that asymptotically large extremal black holes have $\mathcal{Z}=1$, it is natural to suppose that the asymptote $\mathcal{Z}=1$ should be approached from above \cite{ArkaniHamed:2006dz} (see Fig. \ref{fig:HD}). In other words, at fixed charge, higher-derivative corrections should \emph{decrease} an extremal black hole's mass. Moreover, a shift $\Delta\mathcal{Z}>0$ is consistent with the motivation for the WGC from the perspective of extremal black hole decay. For an extremal black hole with charge $Q$ and mass $M$ to be unstable, it must be able to decay into two or more particles whose total charge is $Q$ and total mass is less than $M$. This requires that at least one of the lighter states has $\mathcal{Z}>\frac{Q}{M}$. If $\mathcal{Z}$ approaches unity from above, then a lighter state with larger $\mathcal{Z}$ is simply provided by a smaller extremal black hole.

As was mentioned in \cite{ArkaniHamed:2006dz} and later elaborated upon in \cite{Kats:2006xp,Cheung:2018cwt,Hamada:2018dde,Bellazzini:2019xts,Cheung:2019cwi}, requiring $\Delta\mathcal{Z}>0$ corresponds to a positivity bound on a certain combination of Wilson coefficients of the higher-derivative corrections. For example, consider a subset of 
the leading corrections to the four-dimensional Einstein-Maxwell action (involving only gauge fields)
\begin{align}
{\cal L} =\frac{M_P^2}{2}\mathcal{R} -F_{\mu\nu}^2+a(F^2)^2+b(F\tilde{F})^2
\end{align}
The charge-to-mass-ratio for a large extremal electrically-charged black hole is then shifted by \cite{ArkaniHamed:2006dz}
\begin{align}
\Delta \mathcal{Z}=\frac{a}{5Q^2}+\mathcal{O}(Q^{-3})
\end{align}
Thus for positive $a$ one has $\Delta \mathcal{Z}>0$ and extremal black holes can decay into smaller extremal black holes.

The positivity bound for the four-derivative terms modifying an electrically-charged Reissner-Nordstr\"om black hole was calculated in \cite{Kats:2006xp} and has been shown to be satisfied in 
some (heterotic) string theory examples. 
More recently, it has been shown that under certain assumptions, this positivity bound follows from causality and unitarity constraints on scattering amplitudes \cite{Hamada:2018dde} (see also \cite{Bellazzini:2019xts}). The references \cite{Cano:2019ore,Reall:2019sah} calculated higher-derivative corrections to Kerr black holes respectively up to six derivatives and eight derivatives.

However, while $\alpha'$ corrections to extremal black holes provide us with superextremal states, they do not lead to an entirely satisfying physical picture. The above calculations are valid only for sufficiently large black holes (such that merely including the leading perturbative $\alpha'$ correction is justified). What is the fate of a smaller extremal black hole, for which higher-order $\alpha'$ corrections are important? It is not immediately obvious that these higher-order corrections will also shift $\mathcal{Z}$ positively.

Moreover, the goal of the swampland program is to rule out effective models based on their \emph{low-energy} physics. The masses of these (heavy) black holes are above the cutoff of a low-energy observer. Can the fact that $\Delta \mathcal{Z}>0$ for large extremal black holes give us a useful prediction for the low-energy observer, like the existence of a light state with $\mathcal{Z}>1$? Now that we have learned that $\Delta {\cal Z}>0$ for large extremal black holes in a wide variety of theories, it would be welcome if we could derive a more rigorous statement about the light spectrum. In this note, we take a first step in this direction and provide evidence that in theories with a worldsheet description that is modular invariant, the existence of a light state with $\mathcal{Z}>1$ is related to having a positive charge-to-mass ratio shift for extremal black holes, that is $\Delta \mathcal{Z}>0$.

\section{Modular Invariance and Perturbative String States} \label{sec:MI}
We have seen that satisfaction of the mild WGC by corrections to the large black hole extremality bound does not provide an entirely satisfying picture. We do not learn anything about the fate of small extremal black holes, and as a quantum gravity constraint on low-energy physics this version of the WGC appears somewhat toothless.

In this note, we suggest that in string theory modular invariance plays an important role in resolving these issues. Modular transformation properties of 
conformal field theory (CFT)
partition functions in the presence of conserved currents with quantized charges give rise to an automorphism of the current algebra, a special case of spectral flow \cite{Schwimmer:1986mf}.
This was 
applied by \cite{Montero:2016tif} to the CFT$_2$ dual of AdS$_3$ (and used to define the WGC in three dimensions), and by
\cite{Heidenreich:2016aqi} to 
the worldsheet CFT.
Given a particular state with known charge and mass in the perturbative spectrum, spectral flow allows one to infer the existence of an infinite family of states with known perturbative masses and charges.\footnote{Although in this section we consider perturbative string theory, similar structure also arises in certain F-theory setups. In \cite{Lee:2018urn,Lee:2018spm,Lee:2019xtm} the authors examined the $g_{\rm YM}\to 0$, $M_P$ fixed limit of 6d F-theory compactifications and found the emergence of tensionless heterotic strings. A subset of the excitations of these strings is described by an elliptic genus. In their setup, the quasiperiodicity of Jacobi forms describing the elliptic genus implies a structure similar to that arising from the spectral flow we study in this section. In \cite{Lee:2019tst} it was found that for 4d F-theory compactifications with certain background fluxes turned on, the elliptic genus is not necessarily modular.} 

We now review how spectral flow follows from modular invariance. We point out that spectral flow preserves the extremality of a state with respect to a particular asymptotic charge-to-mass ratio, and can be extrapolated to arbitrarily large mass. It is natural to expect that this perturbative structure has some bearing on black hole physics. We draw a connection between the asymptotic charge-to-mass ratio identified by spectral flow and the asymptotic charge-to-mass ratio of an extremal black hole. Intriguingly, in various setups these two ratios 
agree, suggesting a relationship between superextremal states in both regimes.

Consider the worldsheet CFT of a string theory, whose partition function in the NSNS sector takes the form\footnote{The states we can identify in the perturbative partition function are electrically charged under NSNS sector fields. It would be interesting to study how structure in other sectors of string theory relates to black holes and their higher-derivative corrections.}
\begin{align}\label{eqn:partdef}
	Z(\tau;\mu)\equiv \tr\left(q^{\Delta}y^{Q}\overline{q}^{\tilde{\Delta}}\right)
\end{align}
 where $\tau$ is the modular parameter of the torus, $\mu$ is the chemical potential, $q=e^{2\pi i \tau}$, $y=e^{2\pi i \mu}$, $\Delta=L_0-\frac{c}{24}$, $Q$ is the left-moving charge corresponding to a holomorphic worldsheet current $J$, and tildes represent the right-moving sector. For notational simplicity, we consider a single purely holomorphic current. The straightforward generalization to multiple holomorphic and antiholomorphic currents is presented in \cite{Heidenreich:2016aqi}. Universal modular transformation properties of CFTs \cite{Benjamin:2016fhe} imply that the partition function transforms as
\begin{align}\label{eqn:mod1}
	Z(\tau+1;\mu)=Z(\tau;\mu),\qquad Z(-1/\tau;\mu/\tau)=e^{\pi i k \frac{\mu^2}{\tau}}Z(\tau;\mu)
\end{align}
Here $k$ is the level of the current algebra, with OPE
\begin{align}
	J_L(z)J_L(0)\sim \frac{k}{z^2}+\dots
\end{align}
The level $k$ is non-negative by unitarity, and positive for a non-trivial algebra.

Given a charge lattice $\Gamma_Q$, the dual lattice $\Gamma_Q^*$ is defined by $\Gamma_Q^*=\{\rho~|~\rho Q\in\mathbb{Z}~\forall Q\in\Gamma_Q\}$, which implies that
\begin{align}\label{eqn:charge}
	Z(\tau;\mu+\rho)=Z(\tau;\mu)\qquad \forall \rho\in\Gamma_Q^* ~.
\end{align}
This is merely a $U(1)$ transformation \cite{Montero:2016tif}. Performing a charge transformation (\ref{eqn:charge}) in between two S-transformations (\ref{eqn:mod1}), one has
\begin{align}
	Z(\tau;\mu+\tau\rho)=\exp\left(-\pi i k\left[2\mu\rho+\rho^2\tau\right]\right)Z(\tau;\mu)
\end{align}
Putting in the definition of the partition function (\ref{eqn:partdef}) and pulling the exponential into the trace on the left side, one finds that the spectrum is invariant under the transformation
\begin{align}
	L_0&\to L_0+Q\rho+k\frac{\rho^2}{2}\label{eqn:SF1}\\
	Q&\to Q+k\rho\label{eqn:SF2}
\end{align}
for any $\rho$. This transformation is familiar as a special case of spectral flow \cite{Schwimmer:1986mf}. Thus, given any state in the perturbative spectrum, we can apply spectral flow to the state, generating a tower of new states with different masses and charges.

Consider a string state with perturbative mass $m=\sqrt{\frac{4}{\alpha'}\Delta}=\sqrt{\frac{4}{\alpha'}\tilde{\Delta}}$ and charge $q$. Then, in the large $\rho$ limit, $\mathcal{Z}'^2\equiv\frac{2}{k\alpha'}\frac{q'^2}{m'^2}$ for the transformed state is
\begin{align}
	\mathcal{Z}'^2=1+\frac{k\alpha'}{2}m^2\frac{\left(\mathcal{Z}^2-1\right)}{k^2\rho^2}+\mathcal{O}(\rho^{-3})
\end{align}
\begin{figure}\centering
\includegraphics[width=0.5\textwidth]{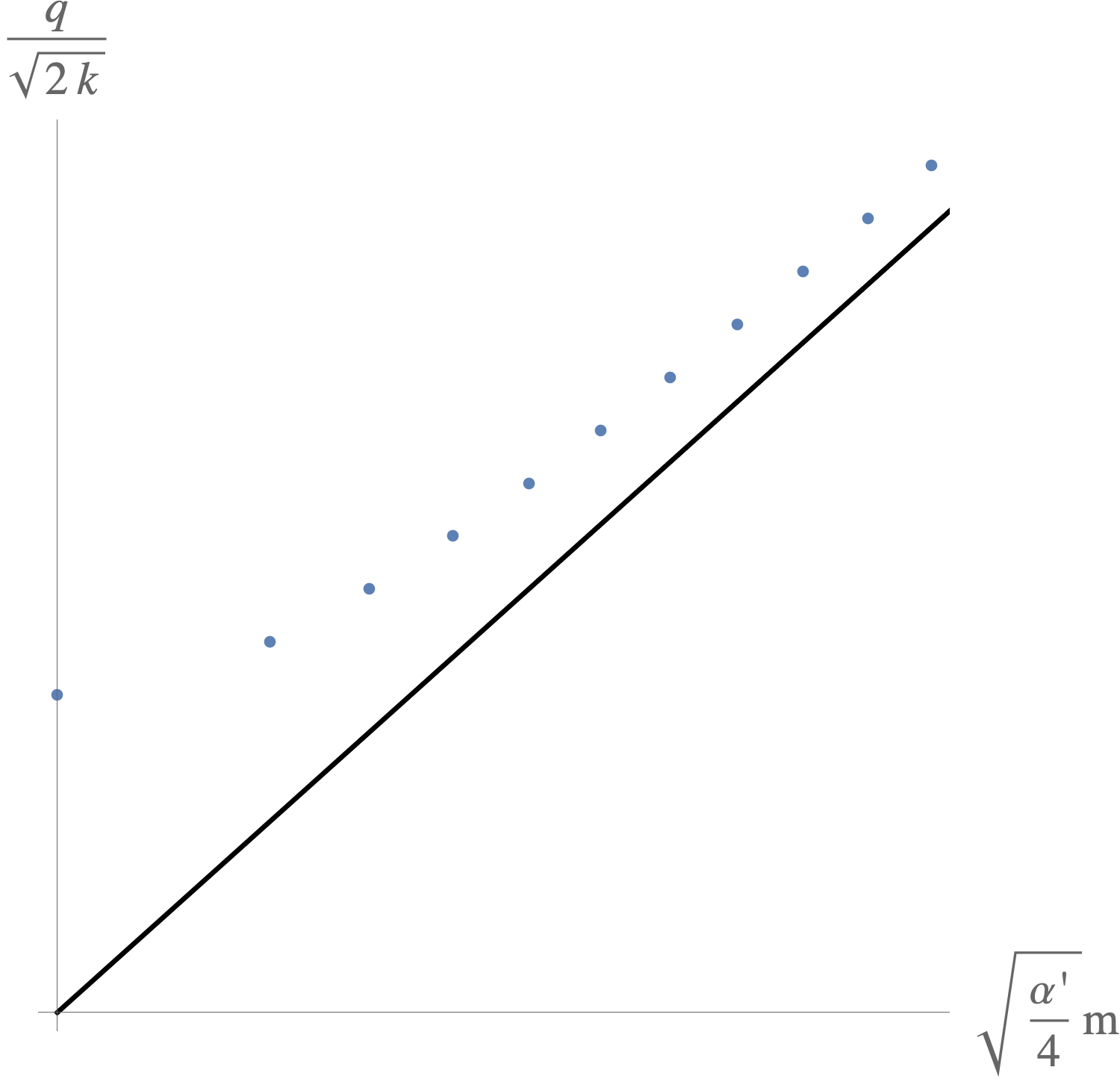}
\caption{Under spectral flow, the sign of $\sqrt{\frac{2}{k\alpha'}}\frac{q}{m}-1$ is preserved. At large mass, string states undergo gravitational collapse and form black holes at small $g_s$.}\label{fig:SF}
\end{figure}
Under spectral flow, states asymptote to $\mathcal{Z}'^2=1$. Moreover, a perturbative state never crosses this asymptotic line (see Fig. \ref{fig:SF}). One may see this by noting that the combination $\Delta-\frac{q^2}{2k}$ is invariant under the transformation (\ref{eqn:SF1}, \ref{eqn:SF2}). In other words, given a state with $\mathcal{Z}> 1$, spectral flow implies the existence of a tower of states monotonically approaching $\mathcal{Z}=1$ from above. Considering the very massive regime, at which a very small but nonzero $g_s$ will cause a state to collapse into a black hole, suggests that $\mathcal{Z}=1$ could perhaps be identified with the asymptotic black hole extremality bound. Indeed, for toroidal orbifolds of type II and heterotic strings, 
the charge-to-mass ratio of the string state approaching the asymptotic line 
matches that of the black hole.
Despite this, one might worry that we do not have a good worldsheet description of black holes generically 
at non-zero coupling,  as in \cite{Heidenreich:2016aqi}. However, as we will see, the symmetries of the near-horizon geometry of an extremal black hole allow us to make more precise the matching of the string and black hole descriptions.
Indeed, we will see in Sec. \ref{sec:BHS} that the entropy of a heterotic two-charge black hole at the transition matches the entropy of a \emph{perturbative} heterotic string, so that at least the perturbative mass relation is sufficiently well-controlled.  Thus quantitative statements can be made about the corrected charge-to-mass ratio at the transition, and we can connect mild and strong forms of the WGC.

This all suggests that the existence of a light state with $\mathcal{Z}>1$ is related to a positive charge-to-mass ratio shift $\Delta \mathcal{Z}>0$ for large extremal black holes. Given a single superextremal perturbative string state, spectral flow dictates that there are very massive superextremal string states, which will collapse into black holes for $g_s\ll 1$. If the correction to the perturbative string charge-to-mass ratio is not too large at the transition, one would expect the corresponding black holes to also have $\mathcal{Z}>1$, so that higher-derivative corrections must shift the extremality bound positively in order to avoid creating a naked singularity. To bound the size of corrections, we need to understand some generalities of the string-black hole transition.

\section{Black Holes as Strings} \label{sec:BHS}

A qualitative description of the string-black hole transition is given by the  ``correspondence principle'' of Horowitz and Polchinski \cite{Horowitz:1996nw}, which formalized and generalized earlier speculations about the relationship between strings and black holes \cite{tHooft:1990fkf,Susskind:1993ws,Sen:1995in}. If we consider a perturbative string state at large level $N\gg 1$ and increase the string coupling slightly, the correspondence principle tells us that at a critical string coupling $g_c\sim N^{-1/4}\ll 1$, the string's Schwarzschild radius will be of the order of the string scale. In $d=4$, interactions will confine the string to the string scale at this coupling\footnote{For subtleties related to spacetime dimension, see \cite{Horowitz:1997jc}.}, and we should begin to regard the state as a black hole.

The critical coupling $g_c$ is determined by setting the curvature (in string units) of the black hole horizon to string scale. Intuitively, at this point $\alpha'$ corrections will start to correct the black hole geometry significantly. Interestingly, one finds that by matching the perturbative string mass and the classical black hole mass at this coupling, the string entropy agrees with the black hole's Bekenstein-Hawking entropy up to factor of order unity for a variety of (even nonsupersymmetric) examples.

We are particularly interested in the corrected charge-to-mass ratio ${\cal Z}$ at the transition. For BPS-saturated states ${\cal Z}$ is protected, so the extremal string charge-to-mass ratio agrees with that of an extremal black hole. For non-BPS states, finite $g_s$ corrections can in general modify the charge-to-mass ratio. However, if the corrections are small enough, a (perturbatively) superextremal string state will remain superextremal at the transition. In this case, the corresponding $\alpha'$-corrected black hole will also have ${\cal Z}>1$, so that higher-derivative corrections must shift the extremality bound positively to avoid a naked singularity.

As a concrete example, we now consider black holes in string theory carrying two electric charges. These black holes have a particularly simple microscopic description as fundamental strings carrying momentum and winding charge.

\subsection{Two-charge black holes}

We now construct four-dimensional black holes carrying winding and momentum charge, following \cite{Horowitz:1997jc}. We start with the metric of a five-dimensional black string in the string frame
\be
ds_{\rm BS_5}^2 =H_1^{-1}\left(-f(r)dt^2+dz^2\right) + f(r)^{-1}dr^2+r^2d\Omega_2^2 ~,
\ee
where
\begin{align}
f(r)&=1-\frac{r_0}{r} ~, \\
H_i &= 1+\frac{r_0}{r}\sinh^2\gamma_i~. \nonumber
\end{align}
Here the $\gamma_i,~i=1,p$ are boost parameters. To add charge in four dimensions, we boost along the $z$ direction
\begin{align}
t &\to \cosh\gamma_p t - \sinh\gamma_p z ~, \\
z &\to \cosh\gamma_p z -\sinh\gamma_p t \nonumber ~.
\end{align}
The boosted black string metric is then
\be \label{eq:BBS}
ds_{\rm BBS_5}^2 = H_1^{-1}\left[-dt^2+dz^2 + \frac{r_0}{r}\left(\cosh\gamma_p dt - \sinh\gamma_p dz\right)^2\right] + f(r)^{-1}dr^2 + r^2d\Omega_2^2 ~.
\ee
We now compactify the $z$ direction $z=z+2\pi R$ and perform a Kaluza-Klein reduction. In string frame, the metric is
\be \label{eq:twochargeBH}
ds^2_{\rm BH_4} = -\frac{f(r)}{H_1H_p}dt^2 + f(r)^{-1}dr^2+r^2d\Omega_2^2 ~,
\ee
In four dimensions, the dilaton is given by
\be
e^{-2\phi} = \sqrt{H_1H_p} ~.
\ee
In Einstein frame, the metric is
\be
ds^2_{\rm BH_4} = -\frac{f(r)}{\sqrt{H_1H_p}}dt^2 + \frac{\sqrt{H_1H_p}}{f(r)}dr^2+r^2d\Omega_2^2 ~
\ee
and the ADM mass is
\be
M_4 = \frac{r_0}{8G_4}\left(2+\cosh 2\gamma_1 + \cosh 2\gamma_p\right) ~,
\ee
where $G_4$ is the four-dimensional Newton constant. The event horizon is located at $r=r_0$ and the two integer-normalized charges corresponding to momentum and winding are given by
\begin{align}
n = \frac{r_0R}{8G_4}\sinh 2\gamma_p ~, \quad w = \frac{r_0\alpha'}{8G_4R}\sinh 2\gamma_1 ~.
\end{align}
These are related to the left and right-moving charges of a fundamental string as
\begin{align}
Q_L &= \frac{r_0}{8G_4}\left(\sinh2\gamma_p - \sinh2\gamma_1 \right) = \frac{n}{R} - \frac{wR}{\alpha'} ~, \\
Q_R &= \frac{r_0}{8G_4}\left(\sinh2\gamma_p + \sinh2\gamma_1 \right) = \frac{n}{R} + \frac{wR}{\alpha'}  ~. \nonumber
\end{align}
We can now take two different extremal limits by
\begin{enumerate}
\item Sending $\gamma_p \to \infty$ and $r_0\to 0$ while keeping $r_0e^{2\gamma_p}$ {and $\gamma_p+\gamma_1$} fixed. This leads to an extremal black hole with
\be
{\cal Z}^2 = \frac{ Q_L^2}{M^2} = 1 ~.
\ee
In heterotic string theory, this black hole is non-BPS.
\item Sending $\gamma_1 \to \infty$ and $r_0\to 0$ while keeping $r_0e^{2\gamma_1}$ {and $\gamma_p-\gamma_1$} fixed. This leads to an extremal black hole with
\be
{\cal Z}^2 = \frac{Q_R^2}{M^2} = 1 ~.
\ee
In heterotic string theory, this black hole is BPS.
\end{enumerate}
The area of the horizon in Einstein frame is given by
\be
A = 4\pi r_0^2\cosh\gamma_1\cosh\gamma_p ~.
\ee
We see that in both extremal limits the area vanishes, so that classically the black holes have vanishing Bekenstein-Hawking entropy. These are thus called small black holes. However, in the extremal limit the horizon size is string scale, so $\alpha'$ corrections should significantly correct the geometry. In the context of heterotic string theory, it is now well understood that $\alpha'$ corrections ``stretch'' the horizon, cloaking the singularity and giving a finite Wald entropy, see \cite{Sen:2007qy} for a review and \cite{Dabholkar:2004yr,Sen:2005kj} for a subset of the original references.\footnote{This conclusion has been contested in \cite{Cano:2018hut} (see also \cite{Cano:2018brq}).}

We now show that including $\alpha'$ corrections gives the black hole a Wald entropy that agrees with the statistical entropy of a perturbative heterotic string. We shall exploit this relation to argue that ${\cal Z}$ receives sufficiently small corrections at the transition to preserve superextremality of a perturbative string state.

\subsection{Higher-derivative corrections, entropy and charge-to-mass ratio}

Typically, extremal black holes have an AdS$_2$ near-horizon geometry. However, in cases when the black hole arises from compactification of a higher-dimensional theory, the AdS$_2$ can combine with an $S^1$ of the compact space to form an AdS$_3$ geometry, or more precisely a BTZ black hole. Examples of this are four-charge black holes in four dimensions and three-charge black holes in five dimensions,
which arise as compactifications of five- and six-dimensional black strings, respectively \cite{Sfetsos:1997xs}. The situation for two-charge black holes is slightly more complicated, since at the classical level their near-horizon geometries are singular and actually correspond to massless BTZ black holes.\footnote{One way of seeing this is by dualizing the fundamental string carrying momentum and winding charges (the F1-P system) to the D1-D5 system \cite{Skenderis:2008qn}.} However, after $\alpha'$ corrections are taken into account, one can have a massive BTZ near-horizon geometry \cite{Castro:2008ne}.

When a near-horizon BTZ geometry exists, the black hole entropy is given by the Cardy formula
\be
S_{\rm BH} = 2\pi\left(\sqrt{\frac{c_L}6h_L} + \sqrt{\frac{c_R}{6}h_R} \right) ~.
\ee
Here, $c_L,c_R$ are the left and right-moving central charges and $h_L,h_R$ are the left and right-moving excitation levels. They are related to the mass and angular momentum of the BTZ geometry via
\be
\ell M_3 =h_L+h_R ~, \quad 8G_3J_3 = h_R-h_L ~,
\ee
where $\ell$ is the AdS$_3$ radius and $G_3$ {(which is small in our semiclassical approximation)} is the three-dimensional Newton constant. The extremality bound is given by
\be
\frac{8G_3|J_3|}{\ell M_3} \leq 1 ~.
\ee
Note that $\ell/G_3 = 2c/3$, where $c$ is the usual Brown-Henneaux central charge \cite{Brown:1986nw}. In terms of the left- and right-moving central charges, we have \cite{Kraus:2005zm}
\be
c = \frac{3\ell}{2G_3} = \frac12\left(c_L+c_R\right) ~.
\ee
The effect of including higher-derivative corrections in the three-dimensional action is to shift the AdS radius as $\ell \to \Omega \ell$, where
\be
\Omega = \frac13g_{\mu\nu}\frac{\partial{\cal L}_3}{\partial R_{\mu\nu}} ~.
\ee
This quantity parametrizes the higher-derivative corrections in terms of the three-dimensional Lagrangian ${\cal L}_3$ \cite{Kraus:2005vz} (see also \cite{Saida:1999ec}). We see that higher-derivative corrections have two effects. First, they modify the effective AdS length and thus the value of the central charges. By Cardy's formula, this implies that the black hole entropy is also corrected. Second, by modifying the AdS length, the extremality bound is also corrected.

For the heterotic two-charge black holes, Kraus and Larsen showed that (under the assumption that a near-horizon BTZ geometry exists after including higher-derivative corrections) the central charges are completely fixed by anomalies \cite{Kraus:2005vz}. The central charges are related to the coefficients appearing in front of Chern-Simons terms. The gravitational Chern-Simons term yields the combination $c_L-c_R$. Supersymmetry in the right-moving sector fixes $c_R$ in terms of the level of the SU(2) current algebra, which can be computed from the SU(2) Chern-Simons term. Because the central charges are fixed by anomalies, they are not modified by other higher-derivative corrections and are therefore exact in $\alpha'$. One thus finds $c_L=24$ and $c_R=12$, so that the entropy of heterotic two-charge black holes is
\be \label{eq:BHentropy}
S_{\rm BH} = 2\pi\left(2\sqrt{h_L} + \sqrt{2h_R} \right) ~.
\ee
This expression exactly matches the statistical entropy of a weakly coupled heterotic string. Recall that the mass of a weakly coupled heterotic string is given by
\be \label{eq:hetstringmass}
M_s^2= \frac{4}{\alpha'}N_R + Q_R^2  = \frac{4}{\alpha'}\left(N_L-1\right) + Q_L^2 ~
\ee
States at $N_R=0$ are BPS-saturated. The entropy of a perturbative string at large excitation level is given by (see e.g. \cite{Halyo:1996xe})
\be
S_{\rm stat} = 2\pi\left(2\sqrt{N_L} + \sqrt{2N_R}\right) ~,
\ee
which matches \eqref{eq:BHentropy} upon identifying $h_{L,R}=N_{L,R}$.\footnote{ Implicitly, we are matching at the string-black hole transition, so this identification holds for $g_s\sim N^{-1/4}\ll 1$.} Notably, this agreement holds for strings with both left- and right-moving sectors excited and therefore even for non-BPS and non-extremal black holes. 

It is remarkable that the black hole entropy matches the entropy of a \emph{perturbative} heterotic string. For an arbitrary (nonsupersymmetric) string state, we might have expected the degeneracy of states to be significantly modified from the perturbative degeneracy at the string-black hole transition, but this is not the case. The appearance of the perturbative string entropy is useful for understanding ${\cal Z}$ at the transition. Since the black hole and perturbative string entropies match at leading order in $g_s$, to this order the charge-to-mass ratio of the black hole is also given by the charge-to-mass ratio of the perturbative string.

{Consider for example the non-BPS state at $N_L=0$ and $N_R\gg 1$. At fixed charge, the mass can be expanded in the closed string coupling $g_s^2$
\begin{align}
	M_s\simeq \sqrt{\frac{4}{\alpha'}N_R}+g_s^2\Delta M(N_R)~.
\end{align}
The mass correction will modify the density of states (at fixed energy) and thus the entropy. Requiring that the entropy is not modified to leading order at the transition means $\Delta M\ll g_s^{-2}\sqrt{N_R/\alpha'}\sim N_R/\sqrt{\alpha'}$, so that in the large $N_R$ limit the mass correction is negligible. In other words, we can trust the perturbative mass relation at the transition for large enough excitation level. This relies crucially on the fact that the entropies match \emph{exactly} to leading order in $g_s$. Given only an approximate entropy matching (as in \cite{Horowitz:1996nw}), the string mass can be corrected by an $O(1)$ factor at the transition. In this case, one might hope to make statements about extremality by generally constraining the sign/magnitude of $\Delta M$ via arguments like self-energy\footnote{For example, gravitational/dilatonic self-interactions should decrease the mass. Moreover, in some cases, the string's ADM mass taking into account backreaction on the massless fields can be calculated \cite{Dabholkar:1997rk}} or explicit loop calculations \cite{Dabholkar:1997fc}.}

For general black holes, the extremality bound is given by
\be \label{eq:BTZextrbound}
\frac{|J_3|}{M_3} = \frac{c}{12}\frac{|N_R-N_L|}{N_R+N_L} \leq \frac{c}{12}~,
\ee
Note that the extremality bound can be modified by $\alpha'$-corrections via the Brown-Henneaux central charge $c=3\ell/(2G_3)$. The extremality bound is saturated for $N_L=0$ or $N_R=0$. In the first case, we have a black hole with $N_R \gg 1$, related to the BTZ charges via
\be
\ell M_3 = \frac{\alpha'}{4}M_s^2 ~, \qquad
8G_3J_3 = \frac{\alpha'}{4}Q_L^2 - 1 ~.
\ee
Thus, the charge-to-mass ratio of this black hole is given by
\be
{\cal Z}_{\rm non-BPS} = \frac{Q_L^2}{M_s^2} = \frac{Q_L^2}{Q_L^2 - 4/\alpha'} > 1 ~.
\ee
In the second case, we obtain an $N_L \gg 1$ BPS-saturated black hole with
\be
\ell M_3 = \frac{\alpha'}{4}M_s^2 + 1~, \qquad
8G_3|J_3| = \frac{\alpha'}{4}Q_R^2 + 1 ~,
\ee
that has a charge-to-mass ratio of
\be
{\cal Z}^2_{\rm BPS} = \frac{Q_R^2}{M_s^2} = 1 ~.
\ee
Thus $\alpha'$ corrections shift the charge-to-mass ratio of the non-BPS extremal black hole positively and do not modify the supersymmetric black hole's charge-to-mass ratio.

For the BPS-saturated state this should not come as a surprise. The BPS bound prevents corrections to the charge-to-mass ratio, guaranteeing the agreement between perturbative strings and black holes. The non-BPS case is more interesting. As we have argued, the matching of entropy implies that we can ignore $g_s$ corrections to the string's perturbative mass formula, and the extremality bound of non-BPS black holes is only modified by $\alpha'$ corrections. Spectral flow on superextremal $(N_L=0)$ string states now guarantees that at large excitation level we will find a black hole with ${\cal Z}>1$. In other words, the existence of black holes with ${\cal Z}>1$ is intimately related to the presence of superextremal states in the light spectrum.

\section{Connection to Positivity of Entropy Corrections} \label{sec:EC}

In \cite{Cheung:2018cwt} it was suggested that the shifts in $\mathcal{Z}_{\rm BH}$ and black hole entropy due to higher-derivative corrections are related. However, the argument of \cite{Cheung:2018cwt} that higher-derivative corrections to Reissner-Nordstr\"om black holes always increase the entropy relies on some assumptions that need not be true \cite{Hamada:2018dde}. Nevertheless, it has been shown for extremal black holes\footnote{In \cite{Hamada:2018dde}, the same result was also shown for general (i.e. nonextremal) electrically-charged Reissner-N\"ordstrom black holes, though an additional assumption about the Gauss-Bonnet term needed to be made.} that an increase in $\mathcal{Z}_{\rm BH}$ due to higher-derivative corrections can be equivalently understood as an increase in entropy \cite{Hamada:2018dde}. The crux of the argument in \cite{Hamada:2018dde} is that the entropy shift for a large extremal black hole (for which higher derivative corrections are small) is dominated by the horizon shift. 
Resolving the degeneracy of the inner and outer horizon for an extremal black hole without introducing a naked singularity requires $\mathcal{Z}_{\rm BH}>1$. 
Here, we briefly mention how the WGC could be related to monotonicity theorems along renormalization group flows for black holes with BTZ near-horizon geometries.

As in Section \ref{sec:BHS}, in some cases a string theory black hole has a BTZ near-horizon geometry. Higher-derivative corrections shift this geometry's Brown-Henneaux central charge, correcting both the entropy and the extremality bound. If the central charge is corrected positively, both the charge-to-mass ratio and entropy are shifted positively. In CFTs, this arises naturally. Consider a CFT$_2$ at a UV fixed point. If we add an irrelevant deformation (such as a higher-derivative correction) to the theory, this triggers a renormalization group flow. If this flow reaches an IR fixed point we have obtained a new IR CFT at which the irrelevant deformation is negligible. The $c$-theorem \cite{Zamolodchikov:1986gt} now implies that the central charge of the theory monotonically decreases along the flow from the UV to the IR. The corresponding black holes with higher-derivative corrections should therefore have larger entropies than their uncorrected counterparts. For the particular class of black holes with BTZ near-horizon geometries, it would be very interesting to confirm this behaviour.

Notice that this is not in conflict with the example of \cite{Hamada:2018dde}, in which a higher-derivative corrected black hole was constructed that has a \emph{lower} entropy than the black hole without higher-derivative corrections. That example violates unitarity, so we do not expect any sensible UV completion admitting a CFT$_2$ dual.

\section{Discussion}\label{sec:DIS}
The WGC has many avatars. In particular, its mild form has been derived via unitarity and causality \cite{Hamada:2018dde} and the AdS/CFT correspondence \cite{Montero:2018fns}. It has also been argued for, though only heuristically, based on the consistency of black hole thermodynamics \cite{Cottrell:2016bty, Cheung:2018cwt}.
Despite this progress, it is not transparent how the mild form of the WGC constrains low-energy physics, which is the goal of the swampland program. Low-energy implications are clearest if we demand the existence of a light superextremal particle, as in some of the stronger forms of the conjecture \cite{ArkaniHamed:2006dz}.

In this note, we have provided evidence that in string theory the existence of a light superextremal particle is deeply related to a positive shift in charge-to-mass ratio for extremal black holes. We observed that under spectral flow (a consequence of modular invariance), the charge-to-mass ratio of an electrically charged perturbative fundamental string asymptotes to unity (in appropriate units) as we increase the perturbative mass. In many cases, this asymptote can be identified as the classical black hole extremality bound. For very massive string states, a small string coupling will suffice to cause a string-black hole transition. To check the preservation of perturbative superextremality for string states, we constructed two-charge black holes in heterotic string theory, microscopically described by fundamental strings with winding and momentum charge. The leading $\alpha'$-correction for these black holes far from extremality was computed in \cite{Giveon:2009da}. Although the leading correction indeed decreases the black hole mass at fixed charge, to make statements about the WGC, we 
need to consider extremal black holes. As these black holes have classically vanishing area, one 
has to sum over infinitely many higher derivative corrections and $\alpha'$-exact techniques would be necessary. 
Using the symmetries of the near horizon geometry of an extremal black hole, we peformed
an $\alpha'$-exact entropy matching to show that in the extremal limit, the $\alpha'$-exact correction to the charge-to-mass ratio of extremal non-BPS two-charge black holes is positive.

This suggests that spectral flow connects having a light superextremal state to the positive shift in charge-to-mass ratio of extremal black holes, at least in the context of heterotic string theory. On the black hole side, it would be interesting to see how this behaviour is encoded in the sign of the coefficients of the infinitely many higher-derivative operators obtained by directly integrating out massive (superextremal) states.

It is worth noting that for very large $k$, the superextremal states related via spectral flow are rather sparse in the charge lattice. Thus, even if one identifies a superextremal black hole, it could be that the lightest superextremal string state is still rather heavy. However, $k$ seems to be weakly bounded above by the central charge of the worldsheet CFT, which is finite in string theory \cite{MiguelWIP}. Thus there does not seem to be a parametric problem.

For the black holes we considered, the microscopics are particularly simple. A natural generalization to consider is four-charge black holes, whose microscopics involve KK-monopoles and NS5 branes \cite{Kutasov:1998zh}. This would more directly connect our results to the works considering higher-derivative corrections to large extremal black holes, such as \cite{Kats:2006xp}, because the extremal four-charge black holes have a finite classical horizon area. 

The connection between corrections to entropy and charge-to-mass ratio observed in \cite{Cheung:2018cwt,Hamada:2018dde} arises naturally for our two-charge black holes, as a consequence of their BTZ near-horizon geometry. For large black holes, this is a consequence of the leading corrections to both the black hole entropy and charge-to-mass ratio being given by the shift of the horizon radius. For black holes with a near-horizon BTZ geometry, this behaviour persists to all orders in $\alpha'$ because both quantities are essentially determined by the Brown-Henneaux central charge. 
As we briefly outlined in Section \ref{sec:EC}, it would be interesting to connect this behaviour to monotonicity theorems along renormalization group flows, which could be used to derive the positivity of corrections to black hole entropy and charge-to-mass ratio for black holes with a near-horizon BTZ geometry.

Finally, another interesting direction of future work would be to not only consider higher-derivative corrections, but also loop corrections, which in string theory scale with $g_s$. For the two-charge black holes we considered these effects are negligible since $g_s \ll 1$, but for sufficiently large (e.g. four-charge) black holes the running of the coefficients of the higher-derivative terms will start to dominate over the higher-derivative operators themselves. In this limit, the running only depends on the massless spectrum of the theory and \emph{increases} the charge-to-mass ratio of nonsupersymmetric extremal black holes \cite{NimaCERN}, vanishing only for BPS-saturated black holes in $N=2$ supergravity \cite{deWit:2010za}. Because the loop corrections are determined by the massless spectrum, they are independent of the UV and are well-known to give rise to universal logarithmic corrections to the black hole entropy. It would be of interest to see if there exists a relation between loop corrections to the charge-to-mass ratio and the entropy of extremal black holes, as is the case for higher-derivative corrections. We hope to come back to some of these questions in future work.

\section*{Acknowledgments}
We thank Alejandra Castro and Ashoke Sen for helpful discussions about small black holes, Pedro Ramirez for bringing the work \cite{Cano:2018hut} to our attention, and Miguel Montero and Toshifumi Noumi for comments on a draft of this paper.
LA would like to thank the University of Wisconsin-Madison for its hospitality, where part of this work was performed. The work of LA is supported by the research program of organization formerly known as the Foundation for Fundamental Research on Matter (FOM), but now as NWO-I, which is part of the Netherlands Organization for Scientific Research (NWO). The work of AC and GS is supported in part by the DOE grant DE-SC0017647 and the Kellett Award of the University of Wisconsin. 

\bibliographystyle{utphys}\bibliography{refs}

\providecommand{\href}[2]{#2}\begingroup\raggedright\begin{thebibliography}{10}

\bibitem{Vafa:2005ui}
C.~Vafa, ``{The String landscape and the swampland},''
\href{http://arxiv.org/abs/hep-th/0509212}{{\ttfamily arXiv:hep-th/0509212
  [hep-th]}}.

\bibitem{Ooguri:2006in}
H.~Ooguri and C.~Vafa, ``{On the Geometry of the String Landscape and the
  Swampland},'' \href{http://dx.doi.org/10.1016/j.nuclphysb.2006.10.033}{{\em
  Nucl. Phys.} {\bfseries B766} (2007) 21--33},
\href{http://arxiv.org/abs/hep-th/0605264}{{\ttfamily arXiv:hep-th/0605264
  [hep-th]}}.

\bibitem{Brennan:2017rbf}
T.~D. Brennan, F.~Carta, and C.~Vafa, ``{The String Landscape, the Swampland,
  and the Missing Corner},'' \href{http://dx.doi.org/10.22323/1.305.0015}{{\em
  PoS} {\bfseries TASI2017} (2017) 015},
\href{http://arxiv.org/abs/1711.00864}{{\ttfamily arXiv:1711.00864 [hep-th]}}.

\bibitem{Palti:2019pca}
E.~Palti, ``{The Swampland: Introduction and Review},''
\newblock 2019.
\newblock
\href{http://arxiv.org/abs/1903.06239}{{\ttfamily arXiv:1903.06239 [hep-th]}}.
\newblock

\bibitem{ArkaniHamed:2006dz}
N.~Arkani-Hamed, L.~Motl, A.~Nicolis, and C.~Vafa, ``{The String landscape,
  black holes and gravity as the weakest force},''
  \href{http://dx.doi.org/10.1088/1126-6708/2007/06/060}{{\em JHEP} {\bfseries
  06} (2007) 060},
\href{http://arxiv.org/abs/hep-th/0601001}{{\ttfamily arXiv:hep-th/0601001
  [hep-th]}}.

\bibitem{Montero:2016tif}
M.~Montero, G.~Shiu, and P.~Soler, ``{The Weak Gravity Conjecture in three
  dimensions},'' \href{http://dx.doi.org/10.1007/JHEP10(2016)159}{{\em JHEP}
  {\bfseries 10} (2016) 159},
\href{http://arxiv.org/abs/1606.08438}{{\ttfamily arXiv:1606.08438 [hep-th]}}.

\bibitem{Heidenreich:2016aqi}
B.~Heidenreich, M.~Reece, and T.~Rudelius, ``{Evidence for a sublattice weak
  gravity conjecture},'' \href{http://dx.doi.org/10.1007/JHEP08(2017)025}{{\em
  JHEP} {\bfseries 08} (2017) 025},
\href{http://arxiv.org/abs/1606.08437}{{\ttfamily arXiv:1606.08437 [hep-th]}}.

\bibitem{Andriolo:2018lvp}
S.~Andriolo, D.~Junghans, T.~Noumi, and G.~Shiu, ``{A Tower Weak Gravity
  Conjecture from Infrared Consistency},''
  \href{http://dx.doi.org/10.1002/prop.201800020}{{\em Fortsch. Phys.}
  {\bfseries 66} no.~5, (2018) 1800020},
\href{http://arxiv.org/abs/1802.04287}{{\ttfamily arXiv:1802.04287 [hep-th]}}.

\bibitem{Kats:2006xp}
Y.~Kats, L.~Motl, and M.~Padi, ``{Higher-order corrections to mass-charge
  relation of extremal black holes},''
  \href{http://dx.doi.org/10.1088/1126-6708/2007/12/068}{{\em JHEP} {\bfseries
  12} (2007) 068},
\href{http://arxiv.org/abs/hep-th/0606100}{{\ttfamily arXiv:hep-th/0606100
  [hep-th]}}.

\bibitem{Cheung:2018cwt}
C.~Cheung, J.~Liu, and G.~N. Remmen, ``{Proof of the Weak Gravity Conjecture
  from Black Hole Entropy},''
  \href{http://dx.doi.org/10.1007/JHEP10(2018)004}{{\em JHEP} {\bfseries 10}
  (2018) 004},
\href{http://arxiv.org/abs/1801.08546}{{\ttfamily arXiv:1801.08546 [hep-th]}}.

\bibitem{Hamada:2018dde}
Y.~Hamada, T.~Noumi, and G.~Shiu, ``{Weak Gravity Conjecture from Unitarity and
  Causality},''
\href{http://arxiv.org/abs/1810.03637}{{\ttfamily arXiv:1810.03637 [hep-th]}}.

\bibitem{Bellazzini:2019xts}
B.~Bellazzini, M.~Lewandowski, and J.~Serra, ``{Amplitudes' Positivity, Weak
  Gravity Conjecture, and Modified Gravity},''
\href{http://arxiv.org/abs/1902.03250}{{\ttfamily arXiv:1902.03250 [hep-th]}}.

\bibitem{Rudelius:2015xta}
T.~Rudelius, ``{Constraints on Axion Inflation from the Weak Gravity
  Conjecture},'' \href{http://dx.doi.org/10.1088/1475-7516/2015/09/020,
  10.1088/1475-7516/2015/9/020}{{\em JCAP} {\bfseries 1509} no.~09, (2015)
  020},
\href{http://arxiv.org/abs/1503.00795}{{\ttfamily arXiv:1503.00795 [hep-th]}}.

\bibitem{Montero:2015ofa}
M.~Montero, A.~M. Uranga, and I.~Valenzuela, ``{Transplanckian axions!?},''
  \href{http://dx.doi.org/10.1007/JHEP08(2015)032}{{\em JHEP} {\bfseries 08}
  (2015) 032},
\href{http://arxiv.org/abs/1503.03886}{{\ttfamily arXiv:1503.03886 [hep-th]}}.

\bibitem{Brown:2015iha}
J.~Brown, W.~Cottrell, G.~Shiu, and P.~Soler, ``{Fencing in the Swampland:
  Quantum Gravity Constraints on Large Field Inflation},''
  \href{http://dx.doi.org/10.1007/JHEP10(2015)023}{{\em JHEP} {\bfseries 10}
  (2015) 023},
\href{http://arxiv.org/abs/1503.04783}{{\ttfamily arXiv:1503.04783 [hep-th]}}.

\bibitem{Brown:2015lia}
J.~Brown, W.~Cottrell, G.~Shiu, and P.~Soler, ``{On Axionic Field Ranges,
  Loopholes and the Weak Gravity Conjecture},''
  \href{http://dx.doi.org/10.1007/JHEP04(2016)017}{{\em JHEP} {\bfseries 04}
  (2016) 017},
\href{http://arxiv.org/abs/1504.00659}{{\ttfamily arXiv:1504.00659 [hep-th]}}.

\bibitem{Lee:2018urn}
S.-J. Lee, W.~Lerche, and T.~Weigand, ``{Tensionless Strings and the Weak
  Gravity Conjecture},''
\href{http://arxiv.org/abs/1808.05958}{{\ttfamily arXiv:1808.05958 [hep-th]}}.

\bibitem{Lee:2018spm}
S.-J. Lee, W.~Lerche, and T.~Weigand, ``{A Stringy Test of the Scalar Weak
  Gravity Conjecture},''
\href{http://arxiv.org/abs/1810.05169}{{\ttfamily arXiv:1810.05169 [hep-th]}}.

\bibitem{Lee:2019xtm}
S.-J. Lee, W.~Lerche, and T.~Weigand, ``{Emergent Strings, Duality and Weak
  Coupling Limits for Two-Form Fields},''
\href{http://arxiv.org/abs/1904.06344}{{\ttfamily arXiv:1904.06344 [hep-th]}}.

\bibitem{Cheung:2019cwi}
C.~Cheung, J.~Liu, and G.~N. Remmen, ``{Entropy Bounds on Effective Field
  Theory from Rotating Dyonic Black Holes},''
\href{http://arxiv.org/abs/1903.09156}{{\ttfamily arXiv:1903.09156 [hep-th]}}.

\bibitem{Cano:2019ore}
P.~A. Cano and A.~Ruipérez, ``{Leading higher-derivative corrections to Kerr
  geometry},'' \href{http://dx.doi.org/10.1007/JHEP05(2019)189}{{\em JHEP}
  {\bfseries 05} (2019) 189},
\href{http://arxiv.org/abs/1901.01315}{{\ttfamily arXiv:1901.01315 [gr-qc]}}.

\bibitem{Reall:2019sah}
H.~S. Reall and J.~E. Santos, ``{Higher derivative corrections to Kerr black
  hole thermodynamics},''
\href{http://arxiv.org/abs/1901.11535}{{\ttfamily arXiv:1901.11535 [hep-th]}}.

\bibitem{Schwimmer:1986mf}
A.~Schwimmer and N.~Seiberg, ``{Comments on the N=2, N=3, N=4 Superconformal
  Algebras in Two-Dimensions},''
\href{http://dx.doi.org/10.1016/0370-2693(87)90566-1}{{\em Phys. Lett.}
  {\bfseries B184} (1987) 191--196}.

\bibitem{Lee:2019tst}
S.-J. Lee, W.~Lerche, and T.~Weigand, ``{Modular Fluxes, Elliptic Genera, and
  Weak Gravity Conjectures in Four Dimensions},''
\href{http://arxiv.org/abs/1901.08065}{{\ttfamily arXiv:1901.08065 [hep-th]}}.

\bibitem{Benjamin:2016fhe}
N.~Benjamin, E.~Dyer, A.~L. Fitzpatrick, and S.~Kachru, ``{Universal Bounds on
  Charged States in 2d CFT and 3d Gravity},''
  \href{http://dx.doi.org/10.1007/JHEP08(2016)041}{{\em JHEP} {\bfseries 08}
  (2016) 041},
\href{http://arxiv.org/abs/1603.09745}{{\ttfamily arXiv:1603.09745 [hep-th]}}.

\bibitem{Horowitz:1996nw}
G.~T. Horowitz and J.~Polchinski, ``{A Correspondence principle for black holes
  and strings},'' \href{http://dx.doi.org/10.1103/PhysRevD.55.6189}{{\em Phys.
  Rev.} {\bfseries D55} (1997) 6189--6197},
\href{http://arxiv.org/abs/hep-th/9612146}{{\ttfamily arXiv:hep-th/9612146
  [hep-th]}}.

\bibitem{tHooft:1990fkf}
G.~'t~Hooft, ``{The black hole interpretation of string theory},''
\href{http://dx.doi.org/10.1016/0550-3213(90)90174-C}{{\em Nucl. Phys.}
  {\bfseries B335} (1990) 138--154}.

\bibitem{Susskind:1993ws}
L.~Susskind, ``{Some speculations about black hole entropy in string theory},''
\href{http://arxiv.org/abs/hep-th/9309145}{{\ttfamily arXiv:hep-th/9309145
  [hep-th]}}.

\bibitem{Sen:1995in}
A.~Sen, ``{Extremal black holes and elementary string states},''
  \href{http://dx.doi.org/10.1142/S0217732395002234}{{\em Mod. Phys. Lett.}
  {\bfseries A10} (1995) 2081--2094},
\href{http://arxiv.org/abs/hep-th/9504147}{{\ttfamily arXiv:hep-th/9504147
  [hep-th]}}.

\bibitem{Horowitz:1997jc}
G.~T. Horowitz and J.~Polchinski, ``{Selfgravitating fundamental strings},''
  \href{http://dx.doi.org/10.1103/PhysRevD.57.2557}{{\em Phys. Rev.} {\bfseries
  D57} (1998) 2557--2563},
\href{http://arxiv.org/abs/hep-th/9707170}{{\ttfamily arXiv:hep-th/9707170
  [hep-th]}}.

\bibitem{Sen:2007qy}
A.~Sen, ``{Black Hole Entropy Function, Attractors and Precision Counting of
  Microstates},'' \href{http://dx.doi.org/10.1007/s10714-008-0626-4}{{\em Gen.
  Rel. Grav.} {\bfseries 40} (2008) 2249--2431},
\href{http://arxiv.org/abs/0708.1270}{{\ttfamily arXiv:0708.1270 [hep-th]}}.

\bibitem{Dabholkar:2004yr}
A.~Dabholkar, ``{Exact counting of black hole microstates},''
  \href{http://dx.doi.org/10.1103/PhysRevLett.94.241301}{{\em Phys. Rev. Lett.}
  {\bfseries 94} (2005) 241301},
\href{http://arxiv.org/abs/hep-th/0409148}{{\ttfamily arXiv:hep-th/0409148
  [hep-th]}}.

\bibitem{Sen:2005kj}
A.~Sen, ``{Stretching the horizon of a higher dimensional small black hole},''
  \href{http://dx.doi.org/10.1088/1126-6708/2005/07/073}{{\em JHEP} {\bfseries
  07} (2005) 073},
\href{http://arxiv.org/abs/hep-th/0505122}{{\ttfamily arXiv:hep-th/0505122
  [hep-th]}}.

\bibitem{Cano:2018hut}
P.~A. Cano, P.~F. Ramírez, and A.~Ruipérez, ``{The small black hole
  illusion},''
\href{http://arxiv.org/abs/1808.10449}{{\ttfamily arXiv:1808.10449 [hep-th]}}.

\bibitem{Cano:2018brq}
P.~A. Cano, S.~Chimento, P.~Meessen, T.~Ortín, P.~F. Ramírez, and A.~Ruipérez,
  ``{Beyond the near-horizon limit: Stringy corrections to Heterotic Black
  Holes},'' \href{http://dx.doi.org/10.1007/JHEP02(2019)192}{{\em JHEP}
  {\bfseries 02} (2019) 192},
\href{http://arxiv.org/abs/1808.03651}{{\ttfamily arXiv:1808.03651 [hep-th]}}.

\bibitem{Sfetsos:1997xs}
K.~Sfetsos and K.~Skenderis, ``{Microscopic derivation of the
  Bekenstein-Hawking entropy formula for nonextremal black holes},''
  \href{http://dx.doi.org/10.1016/S0550-3213(98)00023-6}{{\em Nucl. Phys.}
  {\bfseries B517} (1998) 179--204},
\href{http://arxiv.org/abs/hep-th/9711138}{{\ttfamily arXiv:hep-th/9711138
  [hep-th]}}.

\bibitem{Skenderis:2008qn}
K.~Skenderis and M.~Taylor, ``{The fuzzball proposal for black holes},''
  \href{http://dx.doi.org/10.1016/j.physrep.2008.08.001}{{\em Phys. Rept.}
  {\bfseries 467} (2008) 117--171},
\href{http://arxiv.org/abs/0804.0552}{{\ttfamily arXiv:0804.0552 [hep-th]}}.

\bibitem{Castro:2008ne}
A.~Castro, J.~L. Davis, P.~Kraus, and F.~Larsen, ``{String Theory Effects on
  Five-Dimensional Black Hole Physics},''
  \href{http://dx.doi.org/10.1142/S0217751X08039724}{{\em Int. J. Mod. Phys.}
  {\bfseries A23} (2008) 613--691},
\href{http://arxiv.org/abs/0801.1863}{{\ttfamily arXiv:0801.1863 [hep-th]}}.

\bibitem{Brown:1986nw}
J.~D. Brown and M.~Henneaux, ``{Central Charges in the Canonical Realization of
  Asymptotic Symmetries: An Example from Three-Dimensional Gravity},''
\href{http://dx.doi.org/10.1007/BF01211590}{{\em Commun. Math. Phys.}
  {\bfseries 104} (1986) 207--226}.

\bibitem{Kraus:2005zm}
P.~Kraus and F.~Larsen, ``{Holographic gravitational anomalies},''
  \href{http://dx.doi.org/10.1088/1126-6708/2006/01/022}{{\em JHEP} {\bfseries
  01} (2006) 022},
\href{http://arxiv.org/abs/hep-th/0508218}{{\ttfamily arXiv:hep-th/0508218
  [hep-th]}}.

\bibitem{Kraus:2005vz}
P.~Kraus and F.~Larsen, ``{Microscopic black hole entropy in theories with
  higher derivatives},''
  \href{http://dx.doi.org/10.1088/1126-6708/2005/09/034}{{\em JHEP} {\bfseries
  09} (2005) 034},
\href{http://arxiv.org/abs/hep-th/0506176}{{\ttfamily arXiv:hep-th/0506176
  [hep-th]}}.

\bibitem{Saida:1999ec}
H.~Saida and J.~Soda, ``{Statistical entropy of BTZ black hole in higher
  curvature gravity},''
  \href{http://dx.doi.org/10.1016/S0370-2693(99)01405-7}{{\em Phys. Lett.}
  {\bfseries B471} (2000) 358--366},
\href{http://arxiv.org/abs/gr-qc/9909061}{{\ttfamily arXiv:gr-qc/9909061
  [gr-qc]}}.

\bibitem{Halyo:1996xe}
E.~Halyo, B.~Kol, A.~Rajaraman, and L.~Susskind, ``{Counting Schwarzschild and
  charged black holes},''
  \href{http://dx.doi.org/10.1016/S0370-2693(97)00357-2}{{\em Phys. Lett.}
  {\bfseries B401} (1997) 15--20},
\href{http://arxiv.org/abs/hep-th/9609075}{{\ttfamily arXiv:hep-th/9609075
  [hep-th]}}.

\bibitem{Dabholkar:1997rk}
A.~Dabholkar, ``{Microstates of nonsupersymmetric black holes},''
  \href{http://dx.doi.org/10.1016/S0370-2693(97)00439-5}{{\em Phys. Lett.}
  {\bfseries B402} (1997) 53--58},
\href{http://arxiv.org/abs/hep-th/9702050}{{\ttfamily arXiv:hep-th/9702050
  [hep-th]}}.

\bibitem{Dabholkar:1997fc}
A.~Dabholkar, G.~Mandal, and P.~Ramadevi, ``{Nonrenormalization of mass of some
  nonsupersymmetric string states},''
  \href{http://dx.doi.org/10.1016/S0550-3213(98)00160-6}{{\em Nucl. Phys.}
  {\bfseries B520} (1998) 117--131},
\href{http://arxiv.org/abs/hep-th/9705239}{{\ttfamily arXiv:hep-th/9705239
  [hep-th]}}.

\bibitem{Zamolodchikov:1986gt}
A.~B. Zamolodchikov, ``{Irreversibility of the Flux of the Renormalization
  Group in a 2D Field Theory},'' {\em JETP Lett.} {\bfseries 43} (1986)
  730--732.
[Pisma Zh. Eksp. Teor. Fiz.43,565(1986)].

\bibitem{Montero:2018fns}
M.~Montero, ``{A Holographic Derivation of the Weak Gravity Conjecture},''
  \href{http://dx.doi.org/10.1007/JHEP03(2019)157}{{\em JHEP} {\bfseries 03}
  (2019) 157},
\href{http://arxiv.org/abs/1812.03978}{{\ttfamily arXiv:1812.03978 [hep-th]}}.

\bibitem{Cottrell:2016bty}
G.~Shiu, P.~Soler, and W.~Cottrell, ``{Weak Gravity Conjecture and Extremal
  Black Hole},''
\href{http://arxiv.org/abs/1611.06270}{{\ttfamily arXiv:1611.06270 [hep-th]}}.

\bibitem{Giveon:2009da}
A.~Giveon, D.~Gorbonos, and M.~Stern, ``{Fundamental Strings and Higher
  Derivative Corrections to d-Dimensional Black Holes},''
  \href{http://dx.doi.org/10.1007/JHEP02(2010)012}{{\em JHEP} {\bfseries 02}
  (2010) 012},
\href{http://arxiv.org/abs/0909.5264}{{\ttfamily arXiv:0909.5264 [hep-th]}}.

\bibitem{MiguelWIP}
M.~Montero and G.~Shiu, ``{Work in progress},''.

\bibitem{Kutasov:1998zh}
D.~Kutasov, F.~Larsen, and R.~G. Leigh, ``{String theory in magnetic monopole
  backgrounds},'' \href{http://dx.doi.org/10.1016/S0550-3213(99)00144-3}{{\em
  Nucl. Phys.} {\bfseries B550} (1999) 183--213},
\href{http://arxiv.org/abs/hep-th/9812027}{{\ttfamily arXiv:hep-th/9812027
  [hep-th]}}.

\bibitem{NimaCERN}
N.~Arkani-Hamed, ``Positive geometry of effective field theory.'' Cern winter
  school on supergravity, strings and gauge theory, 2019.

\bibitem{deWit:2010za}
B.~de~Wit, S.~Katmadas, and M.~van Zalk, ``{New supersymmetric
  higher-derivative couplings: Full N=2 superspace does not count!},''
  \href{http://dx.doi.org/10.1007/JHEP01(2011)007}{{\em JHEP} {\bfseries 01}
  (2011) 007},
\href{http://arxiv.org/abs/1010.2150}{{\ttfamily arXiv:1010.2150 [hep-th]}}.

\end{thebibliography}\endgroup

\end{document}